
\input harvmac
\input epsf

\overfullrule=0pt

\def\G{\ifmmode{\cal G}\else$\cal G$\fi}
\def\K{\ifmmode{\cal K}\else${\cal K}$\fi}
\def\L{\ifmmode{\cal L}\else${\cal L}$\fi}
\def\h{\ifmmode{\cal H}\else${\cal H}$\fi}
\def\Q{\ifmmode{\cal Q}\else${\cal Q}$\fi}
\def\T{\ifmmode{\cal T}\else${\cal T}$\fi}
\def\FI{\ifmmode F_\infty\else $F_\infty$\fi}
\def\tp{\tilde{p}}

\def\gone{\epsfxsize=.7cm\epsfbox{g1.eps}}
\def\gtwoone{\epsfxsize=1cm\epsfbox{g2.1.eps}}
\def\gtwotwo{\epsfxsize=.7cm\epsfbox{g2.2.eps}}
\def\gtwothree{\epsfxsize=.7cm\epsfbox{g2.3.eps}}
\def\gthreeone{\epsfxsize=.7cm\epsfbox{g3.1.eps}}
\def\gthreetwo{\epsfxsize=.6cm\epsfbox{g3.2.eps}}
\def\gthreethree{\epsfxsize=1.2cm\epsfbox{g3.3.eps}}
\def\gthreefour{\epsfxsize=.7cm\epsfbox{g3.4.eps}}
\def\gthreefive{\epsfxsize=1cm\epsfbox{g3.5.eps}}
\def\gthreesix{\epsfxsize=1cm\epsfbox{g3.6.eps}}
\def\gthreeseven{\epsfxsize=.7cm\epsfbox{g3.7.eps}}
\def\gthreeeight{\epsfxsize=.7cm\epsfbox{g3.8.eps}}

\lref\Ginsparg{For an old but good review, see P. Ginsparg,
``Matrix models of 2-d gravity,'' Lectures given at Trieste
Summer School, Trieste, Italy, July 22-25, 1991, hep-th/9112013.}
\lref\me{M. Wexler, ``Low temperature expansion of matrix models,''
March 1993, Princeton preprint PUPT-1384, hep-th/9303146.}
\lref\BrezHik{E. Br\'ezin and S. Hikami, {\it Phys. Lett.} {\bf 283B} (1992)
203; Phys. Lett. {\bf 295B} (1992) 209.}
\lref\BIPZ{E. Br\'ezin, C. Itzykson, G. Parisi, and J.-B. Zuber,
{\it Commun. Math. Phys.} {\bf 59} (1978) 35.}
\lref\Riordan{J. Riordan, {\it Bull. Amer. Math. Soc} {\bf 72} (1966)
110.}
\lref\Erdos{P. Erd\H{o}s and A. R\'enyi, {\it Magyar Tudom\'anyos
Akad\'emia Matematikai Kutat\'o Int\'ezet\'enek K\"ozlem\'enyei}
{\bf 5} (1960) 17.}
\lref\ADFO{J. Ambj\o rn, B. Durhuus, J. Fr\"{o}hlich and
P. Orland, {\it Nucl. Phys.} {\bf B270} [FS16] (1986) 457.}
\lref\BKKM{D.V. Boulatov, V.A. Kazakov, I.K. Kostov and A.A. Migdal,
{\it Nucl. Phys.} {\bf B275} [FS17] (1986) 641.}
\lref\BoulKaz{D. V. Boulatov and V. A. Kazakov, {\it Phys. Lett.}
{\bf 186B} (1987) 379.}
\lref\Kostov{I. Kostov, in {\it Non-perturbative
  aspects of the standard model}: proceedings of the XIXth
  International Seminar on Theoretical Physics, Jaca Huesca, Spain,
  6-11 June 1988, eds.~J.~Abad, M.~Belen Gavela, A.~Gonzalez-Arrago
  (North-Holland, Amsterdam, 1989) 295.}
\lref\GN{D.J. Gross and M.J.Newman, {\it Phys. Lett.}
  {\bf 266B} (1991) 291.}

\Title{\vbox{\baselineskip12pt\hbox{PUPT-1398}
\hbox{hep-th/9305041}
}}
{\vbox{\centerline{Matrix Models on Large Graphs}}}
\centerline{{Mark Wexler}\footnote{$^\dagger$}
{E-mail address: \it{wexler@puhep1.princeton.edu}}}
        \vskip2pt\centerline{\it{Department of Physics}}
        \vskip1pt\centerline{\it{Princeton University}}
        \vskip1pt\centerline{\it{Princeton, NJ 08544 USA}}
\rm
\vskip .5in
\noindent
We consider the spherical limit of multi-matrix models on
regular target graphs,
for instance single or multiple Potts models, or lattices
of arbitrary dimension.  We show, to all orders in the
low temperature expansion, that when the degree of the
target graph $\Delta\to\infty$, the free energy becomes
independent of the target graph, up to simple transformations
of the matter coupling constant.  Furthermore, this universal
free energy contains contributions only from those surfaces
which are made up of ``baby universes'' glued together into
trees, all non-universal and non-tree contributions being
suppressed by inverse powers of $\Delta$.  Each order of
the free energy is put into a simple, algebraic form.

\Date{May 1993}

\newsec{Introduction}

Matrix models \Ginsparg~are an efficient way
to describe random surfaces, noncritical string
theory, and perhaps other systems.
A one-matrix model describes pure (i.e., not coupled
to matter) surfaces and the bosonic string in zero
dimensions.
Its partition function is
\eqn\onemat{
\int D\phi \; e^{-{\rm Tr} V(\phi)}}
where $\phi$ is an $N \times N$ hermitian matrix,
and $V$ is some non-gaussian potential; in the
remainder of this work, it will be chosen
to be
\eqn\potential{
V(\phi) = {1\over2}\phi^2 + {g\over\sqrt{N}} \phi^3.}
Multi-matrix models describe random surfaces coupled
to matter, or a string propagating in some target
space.
Their partition functions are of the form
\eqn\multmat{
\int D\phi_1 \cdots D\phi_\xi \;
e^{-{\rm Tr} \left(\sum_i V(\phi_i) - {1\over2}
\phi_i \Gamma_{ij} \phi_j\right)}}
where the coupling matrix $\Gamma$, which depends
on a matter coupling constant $a$, determines the
matter model coupled to the surface, or the
target space of the string.
In this work,
the free energy will always be defined in the
spherical limit $N\to\infty$.

The matrix $\Gamma$ may be thought of as the
adjacency matrix of a labeled graph \G, which we
shall call the {\it target graph}.
One-matrix models have been solved exactly for a
variety of potentials $V(\phi)$ and for particular
genera of the surface, as well as in the
double scaling limit.
Multi-matrix models have also been solved exactly,
but only when the target graph \G~is a tree, that is,
when it has no cycles.
This excludes a great many matter models (or target
spaces), and in particular all those with central
charge greater than one.

A method to deal with arbitrary target graphs has
recently been proposed \me: a low temperature expansion
of the free energy, in powers of the matter coupling
constant $a$.
This reduces scalar multi-matrix model averages to
contractions of tensor one-matrix model averages.
Analogous to the low temperature expansion for
ordinary lattice models,
in $n$-th order one has some number
of ``blobs'' which are open surfaces of uniform
spin, connected by $n$ edges which join blobs of
unequal spin.

In the present work, we consider regular target
graphs, {\it i.e.,} ones whose every vertex has the
same degree or number of neighbors, $\Delta$.
We show, {\it to all orders in the low temperature
expansion}, that in the limit $\Delta\to\infty$ the
free energy of a multi-matrix model becomes independent
of its target graph and equal to some universal function,
\FI, up to some simple transformations of $a$.
Furthermore, we show that the only surfaces which
contribute to the universal model are blobs joined into trees.
Concretely, non-universal and non-tree terms in the
free energy are suppressed by inverse powers of
$\Delta$.
Some examples of models having regular target graphs, which we treat
in detail, are the $q$-state Potts model, $d$-dimensional
lattice, and $\nu$ copies of any regular model, where
$q, d,$ and $\nu \to \infty$.
Finally, we derive a simple, closed expression for
\FI~in any order.

The $d\to\infty$ limit of random surfaces has been discussed
\refs{\ADFO{--}\BKKM}, in the context of the continuum
$d$-dimensional model with Gaussian interaction for the
embedding.\foot{To be contrasted with the Feynman propagators
of the $\L_d$ model considered here}  Both of these works
argue that that the worldsheet has a tree-like (or
``branched polymer'') structure.  Our results agree on
this basic point.  Beyond that, however,
there are differences.  Whereas we obtain entire open surfaces
glued together into trees, according to \refs{\ADFO{--}\BKKM}~the
Gaussian model is dominated in this limit by worldsheet tubes
connected into trees.  There is no contradiction, because the
models are different.  Indeed, it is encouraging that some
kind of trees emerge in both models.  Furthermore, the
``branched polymers'' of \ADFO~appear at one of the critical
regimes of our universal model (see last section): the critical
point of the ``matter''---it is to be remembered that the Gaussian
model is always critical.  The correspondence between our results
and those of \refs{\ADFO{--}\BKKM}~leads one to suspect that our
results are valid non-perturbatively (beyond the low temperature
expansion).

This article is organized as follows.
In the second section we summarize
the old results on the low temperature expansion,
and explain some new ones,
in particular the explicit combinatorial rules for the
diagrams in the partition function and in the
free energy.
In the third section we prove the assertions
concerning universality and trees, and derive
the exact form of \FI, which involves a sum
over labeled trees.
In the fourth section we go through some combinatorial
manipulations to put each term in the sum into
an algebraic form.
In the final section, we discuss some of the implications
of these results, including the critical behavior
of the universal model, and suggest directions for further
work.

\newsec{The low temperature expansion}

\subsec{The target graph}

If the matrix model represents a bosonic string, the target graph
\G~is the discrete target space of the string, its edges
the ``metric.''  For instance, it might be reasonable to
take $\G = {\cal L}_d$, the infinite $d$-dimensional hypercubic
lattice with periodic boundary conditions
whose central charge is $d$.  If the matrix model
represents a random surface, the vertices of the target
graph are the states of the matter, and the (weighted)
edges are the interactions between the
states.\foot{Strictly speaking, the ``target graph''
here should have adjacency matrix $(1-\Gamma)^{-1}$, the matter
propagator.  For the types of matter which we consider
here---(multiple) Potts models---the two are equivalent up
to a rescaling of the matrix fields and a transformation
$a \to -f(a)$, where $f$ is a monotonically increasing
function in the relevant interval, and $f(a)\approx a$ when
$a \ll 1$.}
For example, if the matter is a $q$-state Potts model,
\G~is $\K_q$, the complete graph on $q$ vertices.  The
central charge is $0, {1\over2}, {4\over5}$, and 1 for
$q=1, 2, 3$, and 4, and undefined for $q>4$ where KPZ
scaling breaks down.
So far, all edges of the target graph have the same weight,
$a$.
In either case, we shall use the word ``spin'' to designate
the value of the vertex of \G~assigned to each plaquette
of the surface.

In the case of a random surface with matter, one might
also consider multiple ``uncoupled'' matter models \BrezHik.
For instance, $\nu$ species of, say, Potts models live on
each elementary polygon of the surface, each spin interacting
with nearest neighbor spins of the {\it same} species only.
On a fixed surface this would be trivial: the partition function
would just factorize.  On a random surface, however, the different
species effectively interact with each other because each
interacts with the surface.
The central charge of a multiple model is the sum of the
individual central charges; the coupling matrix $\Gamma$ is
the direct product of the individual coupling matrices.
As a result, the edges of the target graph of a multiple model
never have uniform weights.
For $\nu$ Ising models, for example, the vertices of \G~are
those of a $\nu$-dimensional hypercube, labeled $0, 1,
\ldots, 2^\nu-1$, but each vertex is connected to every other;
any two vertices are connected with the weight $a^\delta$,
where $\delta$ is the number of digits by which the binary
representations of the labels of the two vertices differ.
We shall call this target graph $\Q_\nu$.

It should be noted that all target graphs mentioned above
are regular.
The degree $\Delta=d$ and $q-1$ for the lattice and Potts
model, respectively.
For multiple models, we define $\Delta$ to be the degree
of $\tilde{\G}$, the graph \G~with all $a^2, a^3$, etc. edges
removed;
for $\nu$ Ising models, therefore, $\tilde{\G}$ is the ordinary
$d$-dimensional hypercube and $\Delta=\nu$.
In general, for $\nu$ copies of a single model with degree $\delta$,
$\Delta=\nu\delta$.
In this work, we shall treat only regular target graphs,
although one would get similar results
if this condition were to be somewhat relaxed.

\subsec{The partition function}

For a particular target graph \G~with $\xi$ vertices,
the partition function of the matrix model is defined
\eqn\zdef{\eqalign{
Z_\G(g,a) & = {\int D\mu(\phi_1) \cdots D\mu(\phi_\xi)\;
e^{{1\over2} {\rm Tr} \phi_i\Gamma_{ij}\phi_j}
\over D\mu(\phi_1) \cdots D\mu(\phi_\xi)}
\equiv \left\langle
e^{{1\over2} {\rm Tr} \phi_i\Gamma_{ij}\phi_j}
\right\rangle \cr
D\mu(\phi) & \equiv D\phi\; e^{-{\rm Tr} V(\phi)} \cr}}
The coupling matrix $\Gamma$ depends on the matter coupling
constant $a$ ($a=e^{-\beta}$, where $\beta$ is the inverse
of the matter temperature).
We wish to expand the partition function in powers of $a$:
\eqn\zdeftwo{
Z_\G(g,a) = 1 + \sum_{n=1}^{\infty} {z_n\over n!} a^n}
For single models---where all edges of \G~are weighted
$0$ or $a$---the $n$-th order coefficient $z_n$ is a
sum of possible terms of the form $\langle {\rm Tr}\,
\phi_{i_1}\phi_{j_1} \cdots {\rm Tr}\,\phi_{i_n}\phi_{j_n}
\rangle$, where some of the indices may be equal.
For a multiple model, in $n$-th order one has a sum
of terms $\langle {\rm Tr}\,
\phi_{i_1}\phi_{j_1} \cdots {\rm Tr}\,\phi_{i_m}\phi_{j_m}
\rangle$ where $m \leq n$ and
\eqn\multprod{
\prod_{k=1}^m G_{i_k j_k} = a^n.}
Each term $\langle {\rm Tr}\, \phi_i \phi_j \cdots\rangle$
may be thought of as a subgraph $\h \subseteq \G$: the
vertices of \h~are a subset of those of \G, and the edges
likewise, except that \h~is allowed to have multiple edges.
Each factor ${\rm Tr}\, \phi_i \phi_j$ gives an edge in
\h~between vertices $i$ and $j$.
Many subgraphs are isomorphic and therefore equal, for
example $\langle {\rm Tr}\,\phi_1\phi_2\rangle$ and
$\langle {\rm Tr}\,\phi_2\phi_3\rangle$.  We shall
therefore take the subgraphs \h~to have unlabeled edges
and vertices, because it is much more efficient; but in
counting them, we shall remember that their {\it edges}
are labeled.

Consider, for instance, the term
\def\friend{$\langle {\rm Tr}\, \phi_1\phi_2\,\, ({\rm Tr}\,
\phi_2\phi_3)^2\,\, {\rm Tr}\,\phi_3\phi_1 \rangle$}
\friend, which might arise in fourth or higher order
in the partition function.  Its corresponding (unlabeled)
graph \h~is
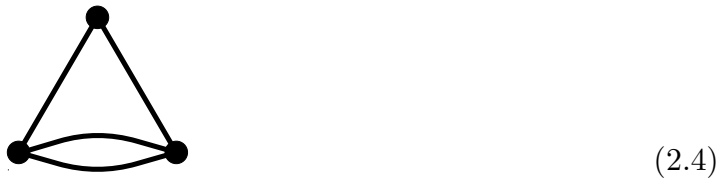
\eqn\friendsk{
\epsfxsize=2cm \epsfbox{friend.eps}
}
The coefficient multiplying this term counts the number
of ways in which the graph \h, with labeled edges, may be embedded
in \G.
In the $q$-state Potts model ($\G=\K_q$), \h~occurs only in $z_4$ with
the coefficient ${1\over2} q(q-1)^2$; in $\nu$ Ising models
($\G=\Q_\nu$), with the coefficient $2^{\nu-1}\nu^2(\nu-1)$
in $z_5$ and with other coefficients in higher orders; and \h~does
not occur in the lattice $\L_d$.

The interpretation in terms of surfaces is as follows: each
vertex of a subgraph \h~is an open surface, a ``blob.''
All spins on this blob are equal, or ``frozen'' at the same
value, and therefore the surface can be described by a
pure gravity matrix model (see below).  Spins on two neighboring
blobs are unequal.
It should be noted that a blob contains contributions from
both connected and disconnected surfaces.
For instance, one of the graphs\foot{We use the words ``graph''
and ``surface'' interchangeably for two objects that are dual
to one another.}
which contribute to \friendsk~is
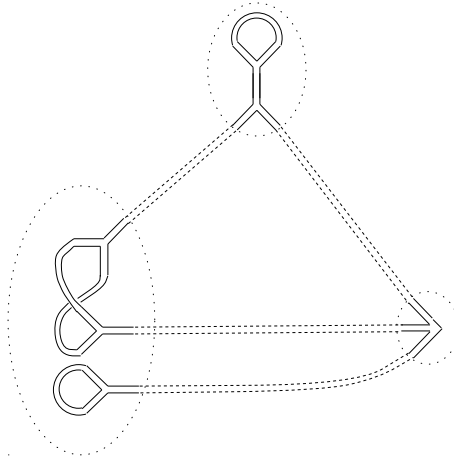
$$
\epsfxsize=5cm \epsfbox{friend2.eps}
$$
Each blob has many internal edges which connect equal spins.
There are only four edges (``bridges,'' dotted in the drawing)
which connect blobs of unequal
spin.
We shall call the subgraphs \h~``Z-skeletons''; the meat on
each vertex of a Z-skeleton is a blob, an open surface
with uniform spin.

As a concrete example, we give the low temperature expansions
of the partition functions to third order for three different
models.
The Z-skeletons are represented graphically, as above.
First, the partition function of the $q$-state Potts model,
$\G = \K_q$:
\eqn\zpotts{\eqalign{
Z & = 1 + {q\choose2} \gone \; a \cr
&\qquad + {1\over2} \left[ {q\choose2} \gtwotwo
+ 6 {q\choose3} \gtwoone
+ 6 {q\choose4} \gtwothree \right] a^2 \cr
&\qquad + {1\over6} \Biggl[ {q\choose2} \gthreeone
+ 24 {q\choose4} \gthreetwo
+ 72 {q\choose4} \gthreethree
+ 6 {q\choose3} \gthreefour \cr
&\qquad\qquad + 18 {q\choose3} \gthreefive +
180 {q\choose5} \gthreesix +
18 {q\choose4} \gthreeseven \cr
&\qquad\qquad + \left({q\choose2}^3 - {q\over4}(q-1)
(6q^3-35q^2+77q-60)\right) \gthreeeight\Biggr] a^3 \cr
&\qquad + {\cal O}(a^4)\cr}}
\vfill
Now a $d$-dimensional hypercubic lattice, $\G = \L_d$
with length L (which will drop out
later in the calculations) and periodic boundary conditions
($\ell \equiv d L^d$):
\eqn\zlat{\eqalign{
Z & = 1 + \ell\; \gone \; a \cr
&\qquad + {1\over2} \left[\ell\; \gtwotwo
+ 2(2d-1)\ell\; \gtwoone
+ (\ell-4d+1)\ell\;\; \gtwothree \right] a^2 \cr
&\qquad + {1\over6} \Biggl[\ell\; \gthreeone
+ 2(2d-1)(2d-2)\ell \gthreetwo
+ 6(2d-1)^2 \ell\; \gthreethree \cr
&\qquad\qquad + 6(2d-1)\ell\; \gthreefive +
6(2d-1)(\ell-6d+2)\ell\;\; \gthreesix +
3(\ell-4d+1)\ell\;\; \gthreeseven \cr
&\qquad\qquad +
(\ell^2-12d \ell + 3\ell +40d^2-24d+4)\ell\;\, \gthreeeight\Biggr] a^3 \cr
&\qquad + {\cal O}(a^4)\cr}}
Finally, a multiple model: $\nu$ Ising models, $\G = \Q_\nu$:
\eqn\zising{\eqalign{
Z & = 1 + {\nu\over2} 2^\nu \;\gone \; a \cr
&\qquad + {1\over2} \left[ {\nu\over2} \;\gtwotwo
+ 2\nu(\nu-1) \;\gtwoone
+ {\nu\over2}({\nu\over2}2^\nu-2\nu+1) \;\gtwothree
+ {\nu\over2}(\nu-1) \;\gone\right] 2^\nu a^2 \cr
&\qquad + {1\over6} \Biggl[{\nu\over2} \;\gthreeone
+ \nu(\nu-1)(\nu-2) \;\gthreetwo
+ 3\nu(\nu-1)^2 \;\gthreethree
+ 3\nu(\nu-1) \;\gthreefive \cr
&\qquad\qquad {3\nu\over2}(\nu-1)(2^\nu \nu-6\nu+4) \;\gthreesix +
{3\nu\over4}(\nu-1)(2^\nu \nu-4\nu+2) \;\gthreeseven \cr
&\qquad\qquad + {\nu\over8}
(2^{2\nu}\nu^2 - 12 2^\nu\nu^2 + 6 2^\nu \nu + 40\nu^2
-48\nu+16) \;\gthreeeight \cr
&\qquad\qquad + 3\nu^2(\nu-1) \;\gtwoone +
{3\nu^2\over4}(\nu-1)(2^\nu-4) \;\gtwothree +
{\nu\over2}(\nu-1)(\nu-2) \;\gone \Biggr] a^3 \cr
&\qquad + {\cal O}(a^4)\cr}}

\subsec{Calculating the blobs}

How can we calculate the contribution of each Z-skeleton to
the partition function?
This can be done by expressing multi-matrix averages of traces
in terms of matrix components, so that they can be decomposed
into contractions of one-matrix tensor averages.
For example, our friend \friend~can be decomposed into
\eqn\decomposed{
\langle \phi^\alpha_\beta \phi^\gamma_\delta \rangle
\langle \phi^\beta_\alpha \phi^\epsilon_\zeta
        \phi^\eta_\theta \rangle
\langle \phi^\zeta_\epsilon \phi^\theta_\eta
	\phi^\delta_\gamma \rangle}
In the latter expression, the averages are with respect
to a ``pure gravity'' one-matrix model:
$\langle{\cal O}\rangle \equiv \int D\mu(\phi) {\cal O}(\phi)
/\int D\mu(\phi)$.
The three tensors in this expressions are just the three
blobs of the Z-skeleton
\epsfxsize=4mm \epsfbox{friend.eps}.
Thus each blob with $\delta$ adjoining bridges is a rank
$(\delta,\delta)$ tensor, and each bridge contracts an
upper and a lower index.\foot{It is necessary to
distinguish between upper and lower indices to preserve
the symmetry under unitary transformations.}

The task is to calculate the one-matrix model vertex factor
arising from a $\delta$-valent blob:
\eqn\vertfact{
B^{\alpha_1\cdots\alpha_\delta}_{\beta_1\cdots\beta_\delta}
\equiv \langle \phi^{\alpha_1}_{\beta_1} \;\cdots
\phi^{\alpha_\delta}_{\beta_\delta} \rangle}
A convenient way to do this is to contract \vertfact~with
$\Lambda^{\alpha_1}_{\beta_1} \; \cdots
\Lambda^{\alpha_\delta}_{\beta_\delta}$, where $\Lambda$
is some $N \times N$ hermitian matrix.
Then the right-hand side of \vertfact~becomes
$\langle ({\rm Tr}\;\Lambda\phi)^\delta \rangle$, for which
the generating functional is the external field integral
\eqn\defq{
Q(\Lambda,g) = \int D\mu(\phi) e^{\sqrt{N}\,{\rm Tr}\,\Lambda\phi}}
$P = {1\over N^2} \log Q$ has been explicitly
calculated in the spherical limit
for an arbitrary matrix $\Lambda$
by Kazakov and Kostov \Kostov~and by Gross and
Newman \GN.
See \me~for a summary of the result, as well as for explicit
details of the vertex factor calculation.

To understand the expansion of $Q$ in powers of $\Lambda$,
we put $\Lambda \to \epsilon\Lambda$ and expand $P$:
\eqn\pexp{
P = p_0 - p_1 \epsilon + {1\over 2} p_2 \epsilon^2 -
{1\over 6} p_3 \epsilon^3 + \cdots}
The first few terms are:
\eqn\pexample{\eqalign{
p_1 & = N^{-1} {\rm Tr}\Lambda\; (3g + 108g^3 + 7776g^5 + \cdots)\cr
&\equiv N^{-1} {\rm Tr}\Lambda\; p_{1,1} \cr
p_2 & = N^{-1} {\rm Tr}\Lambda^2\; (27g^2 + 1944g^4 + \cdots) \cr
&\qquad + N^{-2} ({\rm Tr}\Lambda)^2\; (9g^2 + 1296g^4 + \cdots)\cr
&\equiv N^{-1} {\rm Tr}\Lambda^2\; p_{2,1} +
          N^{-2} ({\rm Tr}\Lambda)^2\; p_{2,2} \cr
p_3 & = N^{-1} {\rm Tr}\Lambda^3\; (6g + 540g^3 + 58320g^5 + \cdots) \cr
&\qquad + N^{-2} {\rm Tr}\Lambda\;{\rm Tr}\Lambda^2\;
			(324g^3 + 69984g^5 + \cdots) \cr
&\qquad + N^{-3} ({\rm Tr}\Lambda)^3\; (7776g^5 + \cdots)\cr
&\equiv N^{-1} {\rm Tr}\Lambda^3\; p_{3,1} +
	  N^{-2} {\rm Tr}\Lambda\;{\rm Tr}\Lambda^2\; p_{3,2} +
	  N^{-3} ({\rm Tr}\Lambda)^3\; p_{3,3}\cr}}
We observe that $p_{n,1}/(n-1)!$, the coefficient of $N\,{\rm Tr}\,
\Lambda^n$ in $p_n$ is just the connected $n$-point Green's function
of the cubic one-matrix model \BIPZ, which is as it should be.
What, then, are the other terms which occur: $p_{n,2}$, etc.?
The matrix $\Lambda$ is attached to the external vertices,
so one way to get, say, $({\rm Tr}\Lambda)^2$ is through a
disconnected two-point function.
But we know that $P$ must be connected.
The only other possibility is to have {\it twisted} graphs.
For instance, a typical graph in $p_{2,1}$ is
\eqn\ptwoone{
\epsfxsize=3cm \epsfbox{twist0.eps}
\quad\sim\quad {\rm Tr}\Lambda^2}
while a typical graph in $p_{2,2}$ is
\eqn\ptwotwo{
\epsfxsize=3cm \epsfbox{twist1.eps}
\quad\sim\quad ({\rm Tr}\Lambda)^2}
Both of these graphs are ``spherical,'' because by attaching
tadpole endcaps to the external lines one obtains closed
spherical graphs.
But by tracing the indices in graph \ptwotwo, one finds that
the two $\Lambda$'s never connect, giving the factor
$({\rm Tr}\Lambda)^2$.
The same holds for all $p_{n,\tau>1}$: the surfaces which contribute
to them can be transformed into ordinary connected Green's
functions by untwisting the ends.

The $n$-th order term in $Q$ has the same contributions
as its counterpart in $P$, the twisted and untwisted connected
graphs.
It also has, of course, twisted and untwisted disconnected
graphs.

\subsec{The free energy}

The partition function looks rather complicated: it contains
connected and disconnected Z-skeletons, which contain
connected and disconnected, twisted and untwisted
blobs.  If some of the blobs
are disconnected, the whole graph may still be connected;
but if the Z-skeleton is disconnected, the graph will be
as well.
Things are simplified immensely when one takes the logarithm
to calculate the free energy
\eqn\fdef{\eqalign{
F_\G(g,a) & = {1\over \xi N^2}\log {Z_\G(g,a) \over
Z_\G(g,0)}\cr
& = f_1 a + {1\over2} f_2 a^2 + {1\over6} f_3 a^3 + \cdots\cr}}
(we recall that $\xi$ is the number of vertices (``volume'')
in the target graph \G).
We can further simplify the calculations by expressing
everything in terms of the coefficients $p_{m,\tau}$ in the
connected generating functional $P(\Lambda)$, rather
than those in the actual disconnected blob-generating functional
$Q(\Lambda)$.

When passing from the partition function to the free
energy, one is of course left with only the connected
graphs.
Another major change occurs in the quality of the
skeletons.
Consider, for instance, one Ising model, whose
target graph $\G = \Q_1 = \gone$.  The Z-skeleton \gtwoone~does
not occur in any order, because it is impossible
to embed it in \G.  The Z-skeleton \gtwotwo~however
does occur in second order.  Its bivalent blobs may
be connected or disconnected, which gives rise to
the following possibilities:
\eqn\blobs{
\epsfxsize=1in \epsfbox{blob1.eps} \qquad
\epsfxsize=1in \epsfbox{blob2.eps} \qquad
\epsfxsize=1in \epsfbox{blob3.eps}}
The first of these is disconnected, so it gets canceled.
The last is the skeleton \gtwotwo.
The skeleton in the middle, however, is the
graph \gtwoone, which does not occur as a Z-skeleton.
When we express everything in terms of connected
blobs rather than the ``natural blobs'' which were
first encountered, we find that many new skeletons are
allowed in addition to the Z-skeletons, and that
the combinatorial rules for finding the coefficients
of the skeletons in the free energy are of course
different as well.  We shall call these new skeletons
``F-skeletons.''

First, consider single models.  As a concrete example,
here is the free energy to third order of the $q$-state Potts model:
\eqn\fepotts{\eqalign{
F & = {1\over2} (q-1) \tp_1^2 a +
\left[{1\over2}(q-1)^2\tp_1^2\tp_2 + {1\over4}p_{2,1}^2\right] a^2 +\cr
& \qquad + \left[{1\over2}(q-1)^3\tp_1^2\tp_2^2 +
{1\over6}(q-1)^3\tp_1^3\tp_3 +
{1\over6}(q-1)(q-2)p_{2,2}^3\right. \cr
& \qquad\qquad \left.{1\over2}(q-1)^2\tp_1 p_{2,1} p_{3,1}+
{1\over6}(q-1)^2\tp_1 p_{2,1} p_{3,2} +
{1\over24}(q-1) p_{3,1}^2 \right] a^3 + {\cal O}(a^4)\cr}}
and of the $d$-dimensional lattice ($D\equiv 2d$):
\eqn\felat{\eqalign{
F & = {1\over2} D \tp_1^2 a +
\left({1\over2} D^2 \tp_1^2\tp_2 +
{1\over4} D p_{2,1}^2\right) a^2+\cr
& \qquad + \left({1\over2} D^3 \tp_1^2\tp_2^2 +
{1\over6} D^3 \tp_1^3\tp_3 +
{1\over2} D^2\tp_1 p_{2,1} p_{3,1} +
{1\over6} D^2\tp_1 p_{2,1} p_{3,2} +
{1\over24} D p_{3,1}^2\right) a^3 + {\cal O}(a^4)\cr}}
where $\tp_n = \sum_\tau p_{n,\tau}$.
We would like to identify the above terms with F-skeletons.
This is not hard, because a factor $p_{n,\tau}^k$ means that
there are $k$ $n$-valent vertices in the F-skeleton; $\tau-1$
is the number of twists of the corresponding blob.
This ``degree sequence'' uniquely identifies small graphs,
so that we can make the following dictionary (omitting the
twist index):
$$
\displaylines{
p_1^2 \to \gone \qquad p_1^2 p_2 \to \gtwoone
\qquad p_2^2 \to \gtwotwo
\qquad p_1^2 p_2^2 \to \gthreethree \cr
\qquad p_1^3 p_3 \to \gthreetwo
\qquad p_2^3 \to \gthreefour
\qquad p_1 p_2 p_3 \to \gthreefive
\qquad p_3^2 \to \gthreeone \cr}
$$

We are now ready to give the ``Feynman rules'' for the F-skeletons
of single models.
The F-skeletons \h~are no longer subgraphs of the target graph \G,
as we showed above.
Instead, the vertices of \G~are to be thought of as ``colors''
with which the vertices of \h~are colored---the ordinary
notion of spin.
Two adjacent vertices of \h~must have different colors,
and these two colors must be adjacent in \G.
Thus the target graph \G~provides both the possible
colors for the F-skeletons, as well as the rules on
how the colors may be combined.

For a particular \G, a graph \h~may appear as an F-skeleton
provided that the latter may be colored according to the
rules of the former.
Consider some particular such unlabeled graph \h~with $n$ edges,
and which therefore appears in the $n$-th order term $f_n$.
The coefficient of \h~in $f_n$ is a product of

\item{$\bullet$}the number of distinct ways to label the
edges of \h, which is $n!$ divided by the order of the
automorphism group of \h;

\item{$\bullet$}and a factor $C_\G(\h)$, the number of
ways to color each {\it labeled} version of \h.

Since $f_n$ appears in the free energy divided by $n!$,
and the free energy is divided by $\xi$ (the volume of \G),
the total factor of \h~in the free energy is
\eqn\totsing{
{C_\G(\h) \over \xi\, |{\rm Aut}(\h)|}.}
However, an F-skeleton \h~is represented in the free
energy by more than one combination of the vertex factors
$p_{n,\tau}$ because the blobs may be twisted in different
ways.
For example, \gthreefive~may be represented by
$\tp_1p_{2,1}p_{3,1}$ or by $\tp_1p_{2,2}p_{3,2}$, among
others.
Each of these combinations may have a different coefficient,
for which the expression \totsing~is an {\it upper bound}.
The reason is topological: the whole graph must be
spherical, while certain contractions of twisted blobs may
yield non-spherical graphs.
If one takes the untwisted $n$-valent vertex factors
$p_{n,1}$ for all vertices, one obtains the largest possible
coefficient, which may still be less than the bound
\totsing.\foot{This occurs for F-skeletons with cycles
or multiple edges; see below.}
To illustrate for the F-skeleton \gthreefive~in \fepotts~and
\felat: the upper
bound given by \totsing~is ${1\over2}(q-1)^2$; this is
saturated by $\tp_1p_{2,1}p_{3,1}$; $\tp_1p_{2,1}p_{3,2}$
is suppressed to ${1\over6}(q-1)^2$; the other four
combinations are completely suppressed.

The free energy rules are similar for multiple models,
except that in order $n$ there are F-skeletons with $n$
{\it and fewer} edges, as there are edges in \G~weighted
$a^2$, $a^3$, etc.
For $\nu$ Ising models ($\G=\Q_\nu$), the free energy
to third order is
\eqn\feising{\eqalign{
F & = {1\over2}\nu \tp_1^2 a + \cr
& \qquad + \left[{1\over2}\nu^2\tp_1^2\tp_2 -
{1\over4}\nu p_{2,1}^2 + {1\over4}\nu(\nu+1)\tp_1^2\right] a^2 +\cr
& \qquad + \left[{1\over2}\nu^3\tp_1^2\tp_2^2 +
{1\over6}\nu^3\tp_1^3\tp_3 +
{1\over2}\nu^2\tp_1p_{2,1}p_{3,1} +
{1\over6}\nu^2\tp_1p_{2,1}p_{3,2} +
{1\over24}\nu p_{3,1}^2 + \right.\cr
& \qquad\qquad + \left.{1\over2}\nu^2(\nu-1)\tp_1^2\tp_2 +
{1\over12}\nu(\nu-1)(\nu-2)\tp_1^2\right]a^3 + {\cal O}(a^4)\cr}}
For $\nu=1$ this can be checked \me~against the exact
solution for one Ising model with cubic vertices \BoulKaz.
One uses the same dictionary as before to translate
from symbols to graphs.
In each order $n$, the graphs with $n$ edges follow the
same combinatorial rules as single models.
(To be more precise, the rules that would obtain for
target graph $\tilde{\G}$, the graph \G~with all $a^2,
a^3$, etc. edges deleted.  In the case $\G=\Q_\nu$,
$\tilde{\G}$ is the ordinary $\nu$-dimensional hypercube.)
The graphs with fewer than $n$ edges have modified
coefficients, due to

\item{$\bullet$}a modified coloring factor $C_\G(\h,n)$
(given a vertex of $\Q_\nu$, for example, there are
$\nu$ other vertices reachable by edges weighted $a$,
${\nu\choose2}$ vertices reachable by $a^2$-edges,
and so on);

\item{$\bullet$}an ``weight-entropy'' factor which counts
the number of ways that the weights of the edges can be
chosen so that their weights satisfy \multprod.

In a multiple model, once an F-skeleton appears in $n$-th
order, it appears in all higher orders as well.
The combinatorics of multiple models will be discussed
in greater detail below.

\newsec{Large target graphs}

On examining the expansions \fepotts, \felat, and \feising~for
the free energy of three different models,
one notices some patterns:

\item{(a)}the $n$-th order term $f_n$ is a
degree $n$ polynomial in $\Delta$, the degree of
\G~or $\tilde\G$;

\item{(b)}the coefficient of $\Delta^n c^n$ in
the free energy---which we shall call $h_n$---contains
contributions only from those F-skeletons which
are trees;\foot{This is reminiscent of, but not
identical to, the result of
Erd\H{o}s and R\'enyi \Erdos~that finite random subgraphs
of $\K_n$ for $n\to\infty$ are almost always forests
of trees.}

\item{(c)}for the single models \fepotts~and
\felat, the leading coefficients $h_n$ are,
in fact, identical; the $h_n$ of the multiple
model \feising~are equal to the above if one ignores
all F-skeletons with fewer than $n$ edges;

\item{(d)}the coefficients of the trees do not care
about the twists of the blobs; in other words, the
vertex factors $p_{n,\tau}$ in the tree coefficients
always occur as the sums $\tp_n$.

These patterns are, in fact, not coincidental, and can easily
be shown to hold for all regular target graphs \G, and
to all orders $n$ in the low temperature expansion.
Because of assertion (a) (see below), when $\Delta \to \infty$
we can ignore all terms in order $n$ but $h_n$.
Here we demonstrate the above assertions, and in the
next section we derive simple expressions for $h_n$.

The number of ways to label the edges of an F-skeleton
\h~(which has $v$ vertices) depends only on \h~and
not on the target graph \G.
For single models, the entire dependence of the coefficient
of \h~on \G~is contained in the coloring factor
$C_\G(\h)$;
for multiple models, the dependence on \G~is in
the modified coloring factor $C_\G(\h,n)$.
It is easy to see that $C_\G(\h)$ has the upper bound
\eqn\uppbound{
C_\G(\h) \leq \xi \Delta^{v-1},}
which can be shown by picking an arbitrary vertex
of \h~that can be colored $\xi$ different ways, and
thereafter traveling to every other vertex of \h,
noticing that when one moves from a colored to a
yet-to-be-colored vertex, one has at most $\Delta$
new colors to choose from.
By induction, trees saturate the bound \uppbound.
The argument is analogous for multiple models.
One has $\xi$ colors for the first vertex of \h,
and for each subsequent vertex, traveling along an
$a^k$-edge of \G, one has at most ${\Delta\choose k}
\leq \Delta^k$ colors to choose from;
using \multprod, we obtain the same bound \uppbound.
This demonstrates assertion (a).

By ``tree,'' we mean an F-skeleton which contains neither
cycles nor multiple edges.
Assertion (b) is now trivial, following from the upper
bound \uppbound~and the fact that a tree with $n$ edges
always has $n+1$ vertices, while a connected non-tree
graph with $n$ edges always has $n$ or fewer
vertices.

Having proved (b), we can ignore all graphs but trees
when calculating $h_n$.
Since trees saturate the bound \uppbound, their coloring
factors are asymptotically (for single models)
\eqn\colasymp{
{C_\G(\h)\over\xi} = \Delta^{|\h|} + {\cal O}(\Delta^{|\h|-1})}
Because the coefficient of leading term is unity,
the {\it maximum} contribution of a tree
\h~to $h_n$, from \totsing, is
\eqn\maxcontr{
{1\over|{\rm Aut}(\h)|}}
This maximum is, in fact, always saturated, for
the following simple reason:
one is free to twist any branch of a tree, without
effecting the relative orientation of the other
branches.
In this way, one can untwist any twisted blob:
all twists therefore contribute equally to the
spherical average.
We have demonstrated assertions (c) and (d) for single models,
showing that the coefficient in $h_n$ of any tree F-skeleton
is precisely \colasymp.
Assertion (d) is just as true for multiple models
as it is for single models.
Assertion (c) holds as well, provided that one restricts
oneself to trees with $n$ edges in order $n$,
because when \h~has $n$ edges, $C_\G(\h,n)=C_\G(\h)$
and the weight-entropy factor is one.

As has already been mentioned, all four assertions
are substantiated by the third-order expansions for the
free energy for the Potts model \fepotts, the lattice
\felat, and multiple Ising models \feising.
The only F-skeletons which contribute to $h_3$,
for instance, are the trees \gthreethree~and
\gthreetwo.
Their automorphism groups have order 2 and 6, respectively,
which are the inverses of their coefficients for all
three models.

\newsec{Counting trees}

We have shown that $h_n$, the coefficient of $\Delta^n$
in the $n$-th order term in the free energy,
for any regular target graph describing a single model,
is given by
\eqn\leading{
h_n = \sum_{\T\in T_n} {V(\T)\over|{\rm Aut}(\T)|}}
where $T_n$ is the set of all unlabeled trees with
$n$ edges, and
\eqn\defv{
V(\T) = \tp_1^{v_1} \tp_2^{v_2} \cdots}
where $v_k$ is the number of $k$-valent vertices in the
tree \T.
By putting $a \to b = a\Delta$, we expand $F$ in
inverse powers of $\Delta$:
$F = \FI + {\cal O}(\Delta^{-1})$,
where
$\FI = \sum_{n=1}^{\infty} h_n b^n$.
Given an unlabeled tree \T~with $n$ edges and $n+1$ vertices,
there are $(n+1)!/|{\rm Aut}(\T)|$ distinct ways of labeling its
vertices.  Therefore, \FI~can be expressed in a simpler
way in terms of vertex-labeled trees:
\eqn\finf{
\FI = {1\over b}\sum_{n=2}^{\infty} {b^n\over n!}
\sum_{\T \in S_n} V(\T)}
where $S_n$ is the set of all vertex-labeled trees
with $n$ vertices (of which there are $n^{n-2}$, by Cayley's
theorem).
Once again: \FI---the free energy of the ``universal
model''---is independent of the target graph,
and the same for any single model.

\FI~for multiple models can be expressed just as easily.
First we derive it for $\nu$ Ising models.
A tree F-skeleton \T~with $m$ edges first appears in order
$b^m$, and in every order thereafter.
In $m$-th order, the coefficient of \T~is as for a single
model; in higher order $n$ it is modified, as explained
above.
Specifically, two neighboring points in \T~may have an
edge in \G~that is weighted $a^2, a^3$, etc.
When coloring a new vertex in \T~along an $a^k$-edge,
there are $\nu\choose k$ colors to choose from, as
opposed to $\nu$ colors for single models.
Since we are interested in the leading term in $\nu$,
we can approximate $\nu\choose k$ by $\nu^k/k!$.
The factors $\nu^k$ combine to $\nu^n$ in $n$-th
order (see \multprod).
But there are different combinations of edge weights
which give $\nu^n$: the sum is over all sets of $m$
distinguishable positive integers whose sum is
$n$, where each term is the inverse of the product
of the factorials of the integers.
The reader can work out a few examples for himself,
but it is not hard to see that the extra factor in $n$-th order,
besides the usual $1/|{\rm Aut}(\T)|$ term,
is just the coefficient of $x^n$ in the Taylor
expansion of $(e^x-1)^m$, which is equal to
\eqn\mike{
{1\over n!} \sum_{k=0}^{m-1} (-1)^k {m\choose k} (m-k)^n.}
One obtains the same sum by transforming the coupling
constant $b$ in \finf~
\eqn\btrans{
b \to b' = e^b - 1}
Therefore, up to the transformation \btrans, \FI~for
multiple Ising models is equal to \FI~for
single models.
Note that now \FI~is the coefficient of the leading
power of $\nu$.
The above argument is easily generalizable to $\nu$
copies of any single degree-$\Delta$ model;
the transformation \btrans~becomes
\eqn\gentrans{
b \to b' = e^{b\Delta}-1.}

The sum over trees $\sum_{\T\in S_n} V(\T)$ has been
expressed in an algebraic form by Riordan\Riordan.
One defines the multi-variable Bell polynomials
\eqn\defbell{
Y_n(y_1,\ldots,y_n) = \left. e^{-y(x)}
{d^n \over dx^n} e^{y(x)} \right|_{x=0},
\qquad y(x) = \sum_{k=1}^{\infty}
{y_k\over k!} x^k.}
Riordan then shows that
\eqn\riord{
\sum_{\T\in S_n} V(\T) =
\tp_1^n Y_{n-2}(\theta\tp_2\tp_1^{-1},\ldots,
\theta\tp_{n-1}\tp_1^{-1})}
where---at the end of the calculation---one puts
$\theta^k \to (n)_k = n(n-1)\cdots (n-k+1)$.
For our purposes, this result can be re-expressed
in a more convenient form by noticing that the
Bell polynomial $Y_n$ with the symbol $\theta$
in every argument can be written
\eqn\mybell{
Y_n(\theta y_1,\ldots,\theta y_n) =
\left. {\partial^n\over\partial x^n}
\left[1+y(x)\right]^n \right|_{x=0}}
Therefore the sum over trees reduces to
\eqn\treesum{\eqalign{
\sum_{\T\in S_n} V(\T) & = \left.
{\partial^{n-2}\over\partial \lambda^{n-2}}
\left[{\partial\Pi\over\partial\lambda}\right]^n \right|_{\lambda=0} \cr
\Pi(\lambda,g) & = \tp_1 \lambda + {1\over2} \tp_2 \lambda^2 +
{1\over6} \tp_3 \lambda^3 + \cdots\cr}}

The vertex generating function $\Pi(\lambda)$ is similar
to the external field integral $P(\Lambda)$ (which
generates the twisted vertex factors $p_{n,\tau}$,
see \defq), except that it does not discriminate
the vertex factors according to twist.
Therefore it is given by the much simpler scalar
version of $P(\Lambda)$:
\eqn\defpi{
\Pi(\lambda,g) = {1\over N^2} \log
{\int D\mu(\phi)\,e^{\sqrt{N}\,\lambda{\rm Tr}\phi}\over
\int D\mu(\phi)}}
This may be calculated by the method of orthogonal
polynomials, which gives
\eqn\orth{
\Pi(\lambda,g) = \int_0^1dx\,(1-x)\log {r(x)\over x} -
{1\over N} \log {h(g)\over h(0)}}
where the function $r$ satisfies the equations
\eqn\eqr{\eqalign{
-6 g r & = -\lambda + s + 3 g s^2 \cr
-6 g \lambda & = (1+6 g s)(-\lambda + s + 3 g s^2)\cr}}
and $h(g)=\int_{-\infty}^{\infty}dx \exp\left[\sqrt{N} \lambda x
- V(x)\right]$.
In order $\lambda^3$ and higher, the leading term in $g$
in $\Pi(\lambda,g)$ comes from the second term in \orth,
and in order $\lambda^{n+2}$ is equal to
\eqn\leading{
{(2n)!\over n!(n+2)!} (3g)^n.}

\newsec{Discussion}

We have shown that for regular target graphs of degree $\Delta$
the leading term of the free energy in inverse powers of
$\Delta$, \FI, is identical (for single models) for all
target graphs and contains contributions only from trees.
In particular, after rescaling the matter coupling constant
$a\to b=a\Delta$, the leading term is given simply by
equation \finf.
This, after some combinatorial manipulations and simple
observations concerning trees, can be
shown to be equal to
\eqn\dfe{
\FI = {1\over b}\sum_{n=2}^\infty {b^n\over n!}
\left. {\partial^{n-2}\over
\partial\lambda^{n-2}} \left[{\partial\Pi\over\partial
\lambda}\right]^n
\right|_{\lambda=0}}
where $\Pi(\lambda,g)$ is the one-matrix model average
\defpi.
An almost identical result holds for multiple models.
For $\nu$ copies of a model with target graph \G~with
degree $\Delta$, the leading term in the expansion
of the free energy in inverse powers of $\nu$ is identical
to \dfe~after the transformations $a\to b=a \nu$ and
$b\to b' = e^{b \Delta} -1$.
These results have been shown to hold in every order
of the low temperature expansion.

If $\Delta$ becomes very large---and therefore
the target graph becomes very large---we can ignore all
but the leading term \FI.
For the lattice and multiple 2, 3, or 4-state Potts models,
this limit corresponds to central charge $c\to\infty$.
The random surface picture is that of a series of ``baby
universes'' (blobs) connected to each other as a tree
(F-skeleton).
The spins on each baby universe are frozen equal, while
the spins on neighboring baby universes are unequal.

For ordinary multi-matrix models (on small target graphs),
the cosmological constant $g$ is coupled to the area of
the surface, while the matter coupling constant $a$ is
coupled to the energy of the matter on the surface.
The typical F-skeletons occurring in these models are
highly multiply connected graphs.
There are three critical regimes: adjusting $g$ makes the
area blow up; adjusting $a$ orders the spins;
and adjusting both yields a critical matter model on
a continuous surface.
For large target graphs, when we are left with the universal
model which has only trees, the picture is different.
The cosmological constant is still coupled to the {\it area},
but here this is the area of {\it each} baby universe.
The matter coupling constant, however, is now coupled
to the size (the number of edges, and therefore the number
of vertices) of the tree.
Adjusting $g$ makes the areas of the baby universes blow
up.
Adjusting $a$ makes the size of the tree---the number of
baby universes---blow up.
Finally, adjusting both yields an infinite tree whose
vertices are continuous surfaces, the baby universes.

So one still gets a phase transition by adjusting the
matter coupling constant, only it is a tree-growing---rather
than a matter-ordering---transition.
The matter coupling constant becomes a ``second-quantized''
cosmological constant.
The critical exponent, $\alpha$, of the free energy
is likely to be ${1\over2}$.
The reason is that if one puts the vertex factors
$\tp_n=1$, the sum over trees just reduces to the Cayley
value $n^{n-2}$ in $n$-th order;
the coefficient of $b^n$ grows as $n^{n-2}/n! \approx
n^{-5/2} e^n$, which gives $\alpha = {1\over2}$.
The $n^{-5/2}$ behavior is robust: it seems to be
independent of the values of the vertex factors $\tp_n$;
we therefore expect that $\alpha = {1\over2}$.
It would be interesting to investigate the critical behavior
of \FI~systematically.

Interestingly, $1\over2$ is the value of the exponent found
in \ADFO~and \BKKM, but for the ``first-quantized''
exponent $\gamma_{\rm str}$ rather than the ``second-quantized''
exponent $\alpha$.
For a gaussian model, of course, there is only one coupling
constant and---generically---only one critical regime.
Comparing our results with those of \ADFO, we see that
for the gaussian models one can blow up the trees, but
not the baby universes.
Perhaps this deviation from the universal model indicates
a subtlety in the target space continuum limit.

Finally, we would like to point out the resemblance between
our results and mean field theory.
In ordinary spin models, the free energy becomes universal
(i.e., independent of the type of matter) when the dimension
of the {\it worldsheet} gets large.
Here, the free energy for a two-dimensional worldsheet
is universal when the dimension (or number of nearest
neighbors) of the {\it target graph} is infinite.
In ordinary spin models, the transition to universal,
mean-field behavior occurs abruptly at a finite dimension.
Could this also be the case for random surfaces?

\bigbreak\bigskip\bigskip\centerline{{\bf Acknowledgements}}\nobreak
The author thanks Michael DeWeese, Alexander Migdal,
Alan Sokal, and Joel Spencer for helpful discussions,
and Edouard Br\'ezin for the suggestion to study many
Ising models.

\listrefs
\bye